# Giant Full-Space Anomalous Hall Effect Induced by Non-Coplanar Spin State in Mn-Rich Mn$_3$Sn

Yiming Liu, Xin Liu, Jiayao Zhu, Fengxian Ma, Li Ma,[*] Dewei Zhao, Guoke Li, Congmian Zhen, and Denglu Hou

Hebei Key Laboratory of Photophysics Research and Application, College of Physics, Hebei Normal University, Shijiazhuang 050024, China

Antiferromagnets are promising candidates for next-generation spintronic devices owing to their negligible stray fields and ultrafast spin dynamics. The noncollinear antiferromagnet Mn$_3$Sn exhibits a large anomalous Hall effect (AHE). However, its specific noncollinear spin configuration leads to the forbiddance of the anomalous Hall conductivity from the (0001) basal plane $\sigma_{(0001)}$, limiting practical applications. Here, using first-principles density functional theory, we demonstrate that Mn enrichment in Mn$_3$Sn drives a magnetic transition from the coplanar 120° spin configuration to a non-coplanar state with moments tilted toward the $c$-axis. This transition is primarily mediated by four-spin ring exchange interaction in the local triangular lattice, which breaks the time-reversal symmetry and generates a giant intrinsic anomalous Hall conductivity over the full three-dimensional space in Mn$_3$Sn. The $\sigma_{(0001)}$ as high as ~-468 $\Omega^{-1}$ cm$^{-1}$ and the enhanced $\sigma_{(01\bar{1}0)}$ of ~-229 $\Omega^{-1}$ cm$^{-1}$ are expected in light Mn self-doping of Mn$_3$Sn (Mn$_{3.125}$Sn$_{0.875}$). Unlike previously reported mechanisms relying on external magnetic fields or strain, our approach exploits intrinsic compositional tuning to stabilize a non-coplanar magnetic ground state for realizing a strong full-space AHE in antiferromagnets, providing another viable pathway toward high-performance, low-power spintronic devices.

## I. INTRODUCTION

Antiferromagnets (AFMs) have emerged as compelling platforms for next-generation spintronics, owing to their vanishing stray fields, terahertz-scale dynamics, and robustness against external magnetic perturbations [1-3]. Yet, unlike ferromagnets, AFMs lack a net magnetization, rendering their readout and control inherently challenging [4]. A breakthrough came with the theoretical prediction that certain noncollinear AFMs can host a large anomalous Hall effect (AHE) despite zero net moment, driven not by magnetization but by the Berry curvature of topologically nontrivial bands in momentum space [5, 6]. Subsequent experiments confirm the kagome AFM Mn$_3$Sn exhibits a finite anomalous Hall conductivity (AHC, ~20 Ω$^{-1}$ cm$^{-1}$) at room temperature [7].

---

[*] Contact author. majimei@126.com

As shown in Figure 1a, stoichiometric Mn$_3$Sn has a hexagonal D0$_{19}$ structure (space group *P*6$_3$/*mmc*). In this structure, Mn atoms are arranged in a kagome lattice on the (0001) basal plane along the *c* axis and Sn atoms occupy the centers of the kagome lattice. Below $T_N$ (~420 K), the Mn moments form a noncollinear triangular spin structure, with neighboring magnetic moments aligned at an angle of 120° [8, 9]. The AHE in this system is captured within a framework of multipole moments: whereas bcc Fe derives its Hall response from a dipole moment, the AHE in Mn$_3$Sn arises from a cluster octupole moment formed by the six Mn spins [10, 11]. Crucially, this octupole moment transforms as an axial vector parallel to the (0001) basal plane. Combined with effective time-reversal symmetry $R_S T$ and the mirror symmetry about the (0001) plane, it is forbidden for anomalous Hall conductivity on the basal plane $\sigma_{(0001)}$ [12, 13]. Consequently, all topological responses, including AHE [7], anomalous Nernst effect [14], and magneto-optical Kerr effect [15], are only detectable on crystallographic planes perpendicular to the basal plane [1, 16, 17]. While non-basal thin films could access the AHE, their epitaxial growth demands more stringent growth conditions, stricter lattice matching, and fewer substrate types to choose, which severely limits scalability and compatibility with conventional semiconductor processes [8, 16]. Generally, a buffer layer is necessary to improve the compatibility between the thin film and the substrate [18, 19]. On the contrary, (0001)-oriented Mn$_3$Sn films offer distinct practical advantages in the aforementioned aspects. Thus, the central challenge becomes clear: how to activate a finite anomalous Hall conductivity on the (0001) plane to achieve the full-space topological responses. Existing approaches, including doping, strain engineering [20], and interfacial Dzyaloshinskii-Moriya interaction (DMI) in heterostructures [21, 22], have sought to break the relevant symmetry and realize a non-zero $\sigma_{(0001)}$. Here, we propose a self-doping strategy via substituting Sn with Mn, as shown in Fig. 1b. This approach could locally break the symmetry protection and induce a non-coplanar spin state on the basal plane, thereby generating a non-zero $\sigma_{(0001)}$ [10].

Here, based on first-principles calculations we show that a stable non-coplanar spin state can be constructed by Mn substituting Sn in Mn$_3$Sn through four-spin ring exchange interaction in local triangular lattice. The effect of this spin state evolution on the anomalous Hall effect was quantified. It was found that a strong full-space AHE is expected by slightly Mn self-doping in Mn$_3$Sn (Mn$_{3.125}$Sn$_{0.875}$), the component of the anomalous Hall conductivity $\sigma_{(0001)}$ as high as ~-468 $\Omega^{-1}$ cm$^{-1}$ can be achieved, and the component of the anomalous Hall conductivity $\sigma_{(01\bar{1}0)}$ can be enhanced to ~-229 $\Omega^{-1}$ cm$^{-1}$ simultaneously.

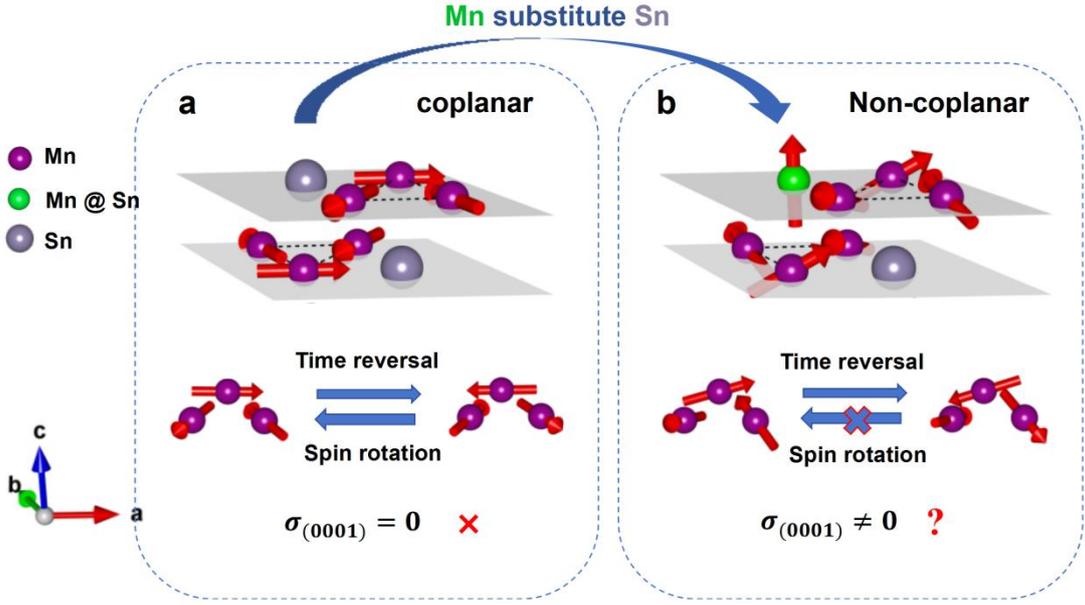

**FIG. 1. Schematics of crystal and magnetic structure for stoichiometric Mn$_3$Sn and Mn-rich Mn$_3$Sn. a.** Coplanar spin structure on the kagome lattice in Mn$_3$Sn. Purple spheres represent Mn atoms, gray spheres represent Sn atoms, and the gray shaded area indicates the basal plane. In this coplanar configuration, time-reversal symmetry $R_S T$ remains a symmetry of the system, leading to a vanishing $\sigma_{(0001)}$. **b.** Non-coplanar magnetic structure in Mn-rich Mn$_3$Sn. Green spheres represent Mn atoms substituting Sn site. The non-coplanarity breaks time-reversal symmetry and can generate a non-zero anomalous Hall conductivity $\sigma_{(0001)} \neq 0$.

## II. METHODS AND COMPUTATIONAL DETAILS

The first-principles calculations were carried out using density functional theory (DFT) [23, 24] utilizing generalized gradient approximation (GGA) [25, 26] for exchange correlation potential. We used Perdew-Burke-Ernzerhof (PBE) functional for the GGA as implemented in Vienna ab initio simulation package (VASP) [27].

Projector-augmented-wave (PAW) potentials were employed to describe electron-ion interactions, setting a planewave-basis cutoff of 550 eV [28]. Starting from the experimentally determined hexagonal unit cell of Mn$_3$Sn ($P6_3/mmc$; $a = b = 5.665$ Å, $c = 4.531$ Å), a 2×1×2 supercell containing 32 atoms was constructed (see Figure 2). All atomic sites in the supercell were systematically labeled for clarity in subsequent analysis. To model Mn-rich compositions, three distinct structural models were generated by substituting Sn with Mn, yielding compositions of Mn$_{24}$Sn$_8$ (nominal Mn$_3$Sn), Mn$_{25}$Sn$_7$ (Mn$_{3.125}$Sn$_{0.875}$), and Mn$_{26}$Sn$_6$ (Mn$_{3.25}$Sn$_{0.75}$). For one Mn doped supercell (Mn$_{25}$Sn$_7$), since all Sn atoms are equivalent to each other, only one configuration needed to be considered by symmetry constraints. For the case of two Mn dopants (Mn$_{26}$Sn$_6$), the most energetically favorable configurations were selected for calculating the AHC. The whole Brillouin-zone was sampled by a 5×10×6 k-mesh for supercells. Lattice constants and atomic coordinates were fully relaxed until the force on each atom was less than 0.001 eV/Å and the total energy minimization was performed with a tolerance of 10$^{-7}$ eV.

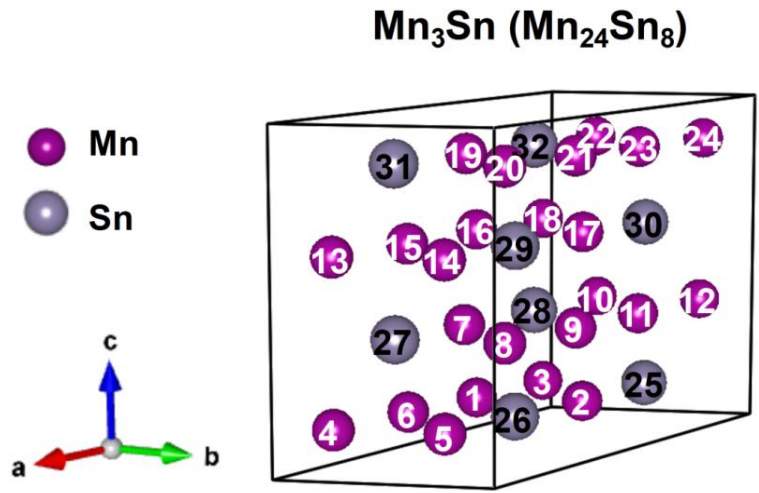

**FIG. 2. Supercell structure of Mn$_3$Sn used in the present study.** Mn and Sn atoms are depicted as purple and gray spheres, respectively. The $a$, $b$, $c$ axes are along the [2$\bar{1}\bar{1}$0], [$\bar{1}$2$\bar{1}$0], and [0001] directions, respectively. Mn atomic sites are labeled with white numbers (1-24), while Sn sites are labeled with black numbers (25-32). The composition Mn$_{25}$Sn$_7$ (corresponding to Mn$_{3.125}$Sn$_{0.875}$) is obtained by substituting the Sn atom at site 27 with Mn; further substitution of the Sn atom at site 32 yields Mn$_{26}$Sn$_6$ (Mn$_{3.25}$Sn$_{0.75}$).

Maximally localized Wannier functions (MLWFs) were utilized to build an accurate tight-binding model via the Wannier interpolation approach [29-31]. Based on these Wannier representations, the intrinsic AHC was computed within the tight-binding framework. The calculations were performed on the constructed tight-binding models using a 101×101×101 k-mesh for all supercells, as implemented in the WannierTools package [32, 33]. The AHC $\sigma_{\alpha\beta}^A$ is obtained from [12, 34]:

$$\sigma_{\alpha\beta}^A = -\frac{e^2}{\hbar}\sum_n \int_{BZ} \frac{d^3k}{(2\pi)^3} f_n(\mathbf{k}) \Omega_{n,\alpha\beta}(\mathbf{k}) , \qquad (1)$$

$$\Omega_{n,\alpha\beta}(\mathbf{k}) = 2i\hbar^2 \sum_{m\neq n} \frac{\langle u_n(\mathbf{k})|\hat{v}_\alpha|u_m(\mathbf{k})\rangle\langle u_m(\mathbf{k})|\hat{v}_\beta|u_n(\mathbf{k})\rangle}{(E_n(\mathbf{k})-E_m(\mathbf{k}))^2} , \qquad (2)$$

where $\Omega_{n,\alpha\beta}(\mathbf{k})$ denotes the Berry curvature of band n in momentum space, $f_n(\mathbf{k})$ is the Fermi–Dirac distribution function, $\hat{v}_{\alpha\,(\beta,\gamma)}$ is the velocity operator, and $E_n(\mathbf{k})$ is the eigenvalue of the *n*th eigenstate $|u_n(\mathbf{k})\rangle$ at **k** point.

## III. RESULTS AND DISCUSSION

### A. Local non-coplanar spin structure in Mn-rich Mn₃Sn

As shown in Figure 3a, the canting angle $\theta$ is defined as the angle between the Mn moment and the basal plane. This parameter quantifies the degree of non-coplanarity in the spin structure: $\theta = 0°$ corresponds to a fully coplanar configuration, where the Mn moments lie entirely within the basal plane. The left panels of Figs. 3b-d show the dependence of the canting angles of Mn moments with the Mn sites in $Mn_3Sn$, $Mn_{3.125}Sn_{0.875}$, and $Mn_{3.25}Sn_{0.75}$ with SOC (yellow bars) and without SOC (green bars), respectively. Only the normal Mn sites labeled 1-24 as shown in Fig. 2 are presented here. The colored dashed circles on the right side of each figure represent the spin structure of the 12 nearest-neighbor Mn atoms surrounding the doping site 27 or 32 in undoped (b) and Mn-doped (c-d) cases.

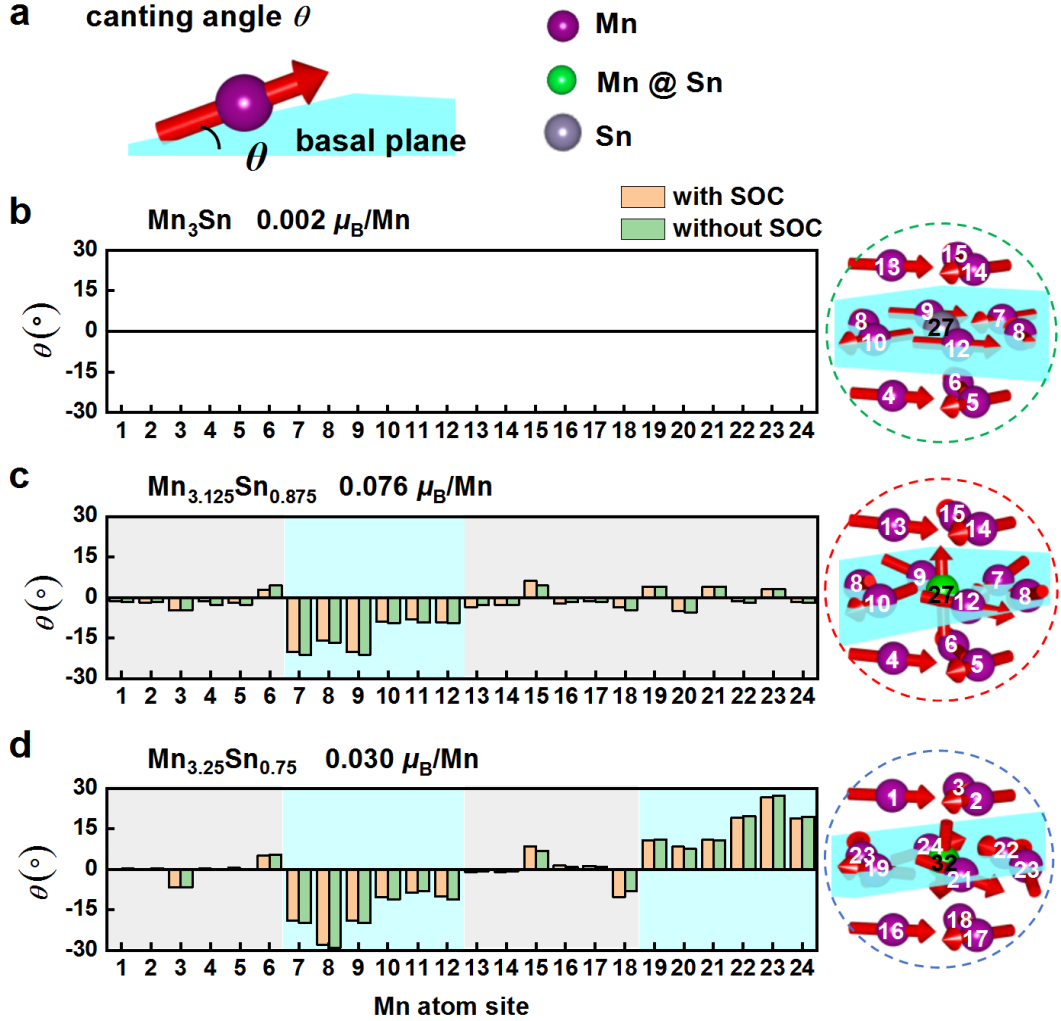

FIG. 3. **Characterization of the canting angles $\theta$ of Mn moments indicative of the non-coplanarity for $Mn_3Sn$, $Mn_{3.125}Sn_{0.875}$, and $Mn_{3.25}Sn_{0.75}$.** **a.** Schematic illustration of the canting angle $\theta$ defined between the Mn moment and the blue basal plane. Canting angles $\theta$ at the 24 regular Mn sites (labeled 1–24, see Figure 2) for **b.** $Mn_3Sn$, **c.** $Mn_{3.125}Sn_{0.875}$, and **d.** $Mn_{3.25}Sn_{0.75}$. Results with and without SOC are shown as yellow and green bars, respectively. Adjacent colored dashed circles on the right side of each panel are the space distribution of spin structures of the 12 nearest-neighbor Mn atoms surrounding the doping site 27 or 32 in b. $Mn_3Sn$, c. $Mn_{3.125}Sn_{0.875}$, and d. $Mn_{3.25}Sn_{0.75}$. Red arrows illustrate both the direction and magnitude of the Mn moments.

In stoichiometric $Mn_3Sn$ (Fig. 3b), all Mn moments remain strictly within the basal plane ($\theta = 0°$), exhibiting a fully coplanar spin structure. The resulting net magnetic moment is negligible (~0.002 $\mu_B$/Mn), in excellent agreement with neutron diffraction experiments [35]. In contrast, the left panel of Fig. 3c reveals that the canting angles of all Mn moments deviate from 0°, indicating that Mn self-doping in Mn$_{3.125}$Sn$_{0.875}$ drives a clear departure from the coplanar spin state. Moreover, the out-of-plane canting is markedly stronger in the blue region compared to the gray region. This spatial correlation demonstrates that the magnitude of the canting angle is directly governed by proximity to the dopant. In other words, the substitutional Mn atom acts as a local perturbation center that reorients neighboring spins within the same kagome layer. Such local symmetry breaking lifts the perfect cancellation of magnetic moments in the ideal coplanar structure, thereby inducing a finite net magnetization of approximately 0.076 $\mu_B$/Mn.

Fig. 3d shows that further increasing the Mn content to Mn$_{3.25}$Sn$_{0.75}$ enhances and stabilizes this non-coplanar character. Two distinct canting regions now emerge around the two Mn dopant sites labeled 27 and 32. Due to antiferromagnetic coupling, the Mn moments surrounding site 27 tilt downward (negative *c*-axis direction), while those near site 32 tilt upward (positive *c*-axis direction). This opposing canting results in partial cancellation of the induced moments, yielding a reduced net magnetization of ~0.030 $\mu_B$/Mn. In summary, Mn substitution at Sn sites not only induces a local non-coplanar spin state but also preserves the overall antiferromagnetic character.

The spin related Hamiltonian for Mn$_3$Sn can be written as [36]:
$$H_S = J_1 \sum_{<ij>_{xy}} \mathbf{S}_i \cdot \mathbf{S}_j + J_2 \sum_{<ij>_z} \mathbf{S}_i \cdot \mathbf{S}_j + \sum_{ij} \mathbf{D}_{ij} \cdot (\mathbf{S}_i \times \mathbf{S}_j) - \sum_i K(\hat{\mathbf{n}}_i \cdot \mathbf{S}_i)^2, \quad (3)$$
where the first and second terms are the in-plane and out-of-plane Heisenberg exchange interactions, respectively; the third term is the DMI; and the fourth term describes the magnetocrystalline anisotropy. Both DMI and magnetocrystalline anisotropy arise from SOC. To determine the origin of the local non-coplanar spin structure in Mn-rich Mn$_3$Sn, we computed the magnetic ground states of Mn$_3$Sn, Mn$_{3.125}$Sn$_{0.875}$, and Mn$_{3.25}$Sn$_{0.75}$ without including SOC. The resulting canting angles $\theta$ shown as green bars in Figs. 3b-d are nearly indistinguishable from those obtained with SOC (yellow bars) for all 24 regular Mn sites (labeled 1-24). This remarkable agreement demonstrates that the non-coplanar spin texture induced by Mn substitution at Sn sites is predominantly governed by Heisenberg exchange interactions, rather than by SOC-driven effects such as DMI or magnetic anisotropy.

In stoichiometric Mn$_3$Sn, the Mn atoms on the basal plane form a global kagome lattice, which constitutes a three-sublattice antiferromagnetic system, as illustrated in Figure 4a. The antiferromagnetic coupling among the three sublattices stabilizes a noncollinear inverse triangular spin structure [7]. The dominant exchange interaction in this system is the conventional two-spin Heisenberg exchange, expressed as [37]:

$$H_{2-\text{spin}} = -\sum_{ij} J_{ij} \mathbf{S_i} \cdot \mathbf{S_j}, \qquad (4)$$

where the exchange integral constant $J_{ij}<0$ reflects antiferromagnetic coupling between Mn spins at sites $i$ and $j$. Upon Mn substitution at Sn sites, a local triangular lattice of Mn atoms emerges around the dopant site, superimposed on the original kagome network, as shown in the area enclosed by the black dashed loop in Fig. 4b. Within this local environment, the substitutional Mn atom and its in-plane neighbors form a four-sublattice antiferromagnetic system. A rhombus-shaped four-site plaquette (labeled $i$-$j$-$k$-$l$) closed loop enables a higher-order, non-Heisenberg magnetic interaction: the four-spin ring exchange, given by [37]:

$$H_{4-\text{spin}} = -\sum_{ijkl} K_{ijkl}[(\mathbf{S_i} \cdot \mathbf{S_j})(\mathbf{S_k} \cdot \mathbf{S_l}) + (\mathbf{S_j} \cdot \mathbf{S_k})(\mathbf{S_l} \cdot \mathbf{S_i}) - (\mathbf{S_i} \cdot \mathbf{S_k})(\mathbf{S_j} \cdot \mathbf{S_l})], (5)$$

where $K_{ijkl}$ is the four-spin exchange integral constant.

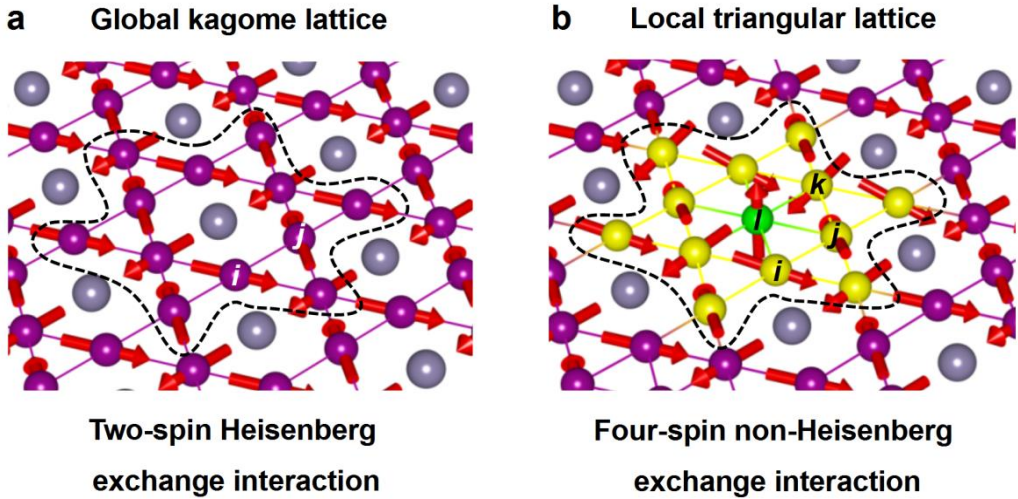

**FIG. 4. The magnetic exchange interactions on the basal plane in stoichiometric and Mn-rich Mn₃Sn. a.** Coplanar 120° antiferromagnetic order in stoichiometric Mn$_3$Sn, stabilized by two-spin Heisenberg exchange interaction within the global kagome lattice. **b.** Local non-coplanar spin structure in Mn-rich Mn$_3$Sn, driven by four-spin ring exchange interaction arising from Mn atoms substituting Sn sites and forming triangular lattice.

Unlike the two-spin Heisenberg interaction, the four-spin ring exchange originates from virtual electron hopping around a closed plaquette and can stabilize complex magnetic structures by permitting a finite solid angle among four spins, thereby enabling non-coplanar spin configurations [38, 39]. Notably, in systems where crystal symmetry forbids the DMI, four-spin exchange alone can stabilize topological spin structures without requiring antisymmetric exchange [40-42]. In Mn-rich $Mn_3Sn$, the interplay between the inherent two-spin Heisenberg exchange of the kagome lattice and the emergent four-spin ring exchange induced by Mn doping gives rise to local non-coplanar spin structures. This competition and cooperation between different exchange pathways provide a microscopic origin for the observed canting angles, even in the absence of strong SOC-driven effects.

### B. Prediction of giant full-space anomalous Hall effect

To further evaluate the AHC of $Mn_3Sn$, $Mn_{3.125}Sn_{0.875}$, and $Mn_{3.25}Sn_{0.75}$, we first performed band structure fitting for each system. As shown in Figs. 5a-c, the band structures obtained from first-principles calculations and those reconstructed via Wannier interpolation are in excellent agreement along the selected high-symmetry path. This consistency not only validates the accuracy of our Wannier fitting approach but also establishes a reliable foundation for subsequent analysis of the topological transport properties in these systems.

We now systematically analyze the AHC of $Mn_3Sn$, $Mn_{3.125}Sn_{0.875}$, and $Mn_{3.25}Sn_{0.75}$, focusing on both the basal plane component $\sigma_{(0001)}$ and the vertical plane component $\sigma_{(01\bar{1}0)}$. Figs. 5d-f display $\sigma_{(0001)}$ as a function of energy, while Figs. 5g-i show $\sigma_{(01\bar{1}0)}$ as a function of energy. For stoichiometric $Mn_3Sn$, $\sigma_{(0001)}$ at the Fermi leve $E_F$ is nearly zero (Fig. 5d), consistent with symmetry constraints and experimental observations [7]. In contrast, Mn enrichment induces a non-coplanar spin structure that breaks time-reversal symmetry. Previous studies have demonstrated that even a slight canting of Mn moments can generate a non-zero $\sigma_{(0001)}$. For instance, Zhu *et al.* reported that under an external magnetic field, when the Mn moments in $Mn_3Sn$ are canted by approximately 3° out of the basal plane, $\sigma_{(0001)}$ becomes non-zero [43]. In our work, Mn enrichment in $Mn_3Sn$ stabilizes a pronounced non-coplanar configuration, resulting in a giant intrinsic $\sigma_{(0001)} \approx -468$ $\Omega^{-1}$ cm$^{-1}$ at $E_F$ for $Mn_{3.125}Sn_{0.875}$ (Fig. 5e). A similar mechanism persists in $Mn_{3.25}Sn_{0.75}$, where $\sigma_{(0001)} \approx -311$ $\Omega^{-1}$ cm$^{-1}$ (Fig. 5f). The $\sigma_{(0001)}$ values significantly exceed the maximum $\sigma_{(0001)}$ (~-140 $\Omega^{-1}$ cm$^{-1}$) at low temperatures reported in $Mn_3Sn$ single crystals [7]. Moreover, they compare favorably with those of other notable noncollinear AFMs such as -284 $\Omega^{-1}$ cm$^-$

[1] in Mn$_3$Rh, -312 $\Omega^{-1}$ cm$^{-1}$ in Mn$_3$Ir, and 98 $\Omega^{-1}$ cm$^{-1}$ in Mn$_3$Pt [12]. Our theoretical calculation may precisely explain the experimental results of Huang *et al.*, where a non-zero AHE was observed in (0001)-oriented Mn-rich Mn$_3$Sn[44]. Similarly, Chatterjee *et al.* observed a non-coplanar spin structure in Mn$_3$Sn/Ta heterostructures with a small canting angle of 0.1-0.2°, leading to a non-zero $\sigma_{(0001)}$ [18].

Figs. 5g-i show the energy-dependent $\sigma_{(01\bar{1}0)}$ for Mn$_3$Sn, Mn$_{3.125}$Sn$_{0.875}$, and Mn$_{3.25}$Sn$_{0.75}$. For Mn$_3$Sn, $\sigma_{(01\bar{1}0)}$ at $E_F$ is approximately -116 $\Omega^{-1}$ cm$^{-1}$, which is consistent with the value reported (~-133 $\Omega^{-1}$ cm$^{-1}$) and the overall shape of the energy-dependent curve [12]. Upon Mn doping, $\sigma_{(01\bar{1}0)}$ at $E_F$ increases significantly, reaching approximately -229 $\Omega^{-1}$ cm$^{-1}$ for Mn$_{3.125}$Sn$_{0.875}$, and -297 $\Omega^{-1}$ cm$^{-1}$ for Mn$_{3.25}$Sn$_{0.75}$, as indicated in Figs. 5h and i, respectively. Notably, these values in Mn-rich Mn$_3$Sn are larger than that of Mn$_3$Ga (~81 $\Omega^{-1}$ cm$^{-1}$) and even approach the value reported for Mn$_3$Ge (~-330 $\Omega^{-1}$ cm$^{-1}$) [12].

The underlying mechanism for the enhanced AHC in both directions originates from a dual effect of Mn enrichment: it causes minor lattice distortions and stabilizes a non-coplanar magnetic structure. This structural and magnetic reconfiguration concurrently breaks spatial inversion and time-reversal symmetries, substantially lowering the overall symmetry of the system. The resulting symmetry breaking affects the electronic band structure and may generate large Berry curvature that does not cancel out. This enables a giant intrinsic AHC within the basal plane, where it was originally vanishing, and simultaneously a giant enhancement of the out-of-plane AHC, thereby achieving a full-space anomalous Hall response. Our results establish self-doping as an effective strategy for engineering a giant, omnidirectional anomalous Hall effect in Mn$_3$Sn.

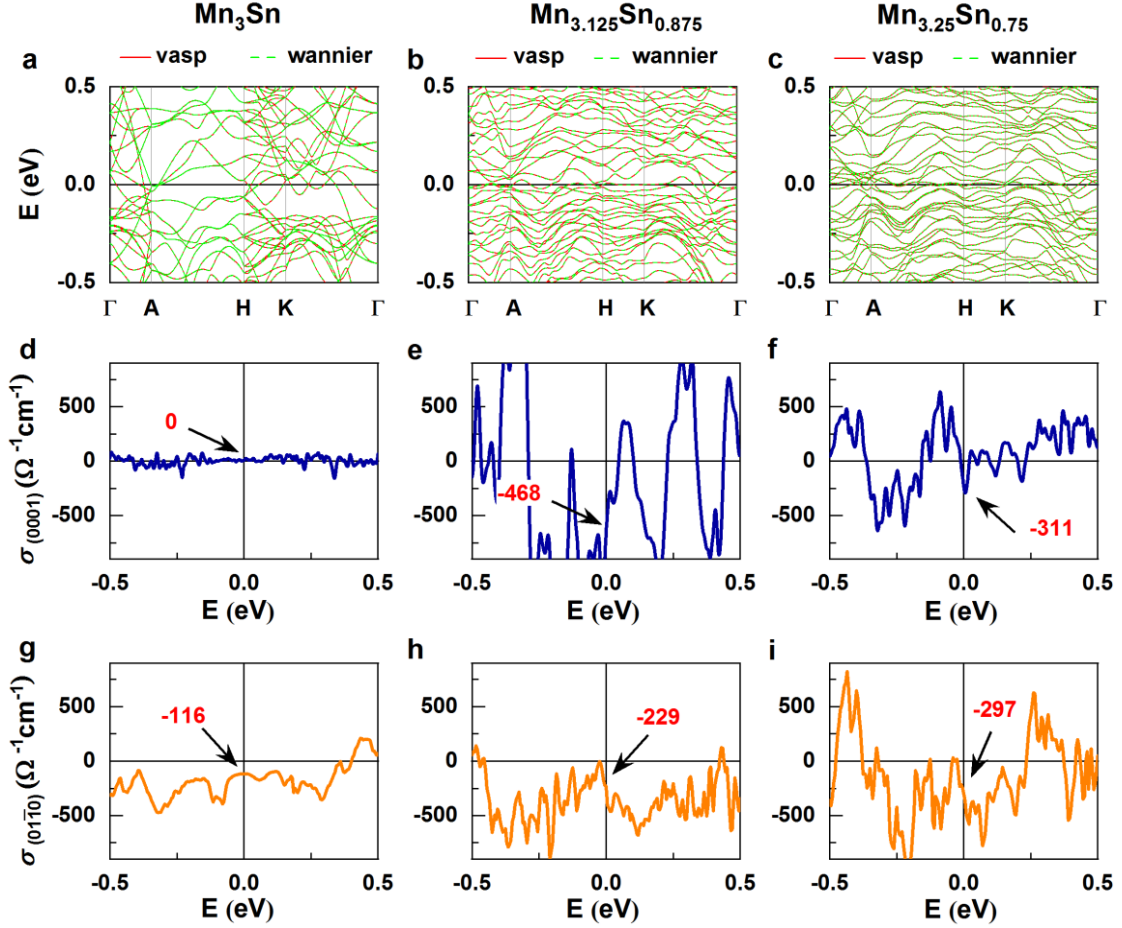

**FIG. 5. Anomalous Hall conductivity (AHC) near the Fermi level for Mn$_3$Sn, Mn$_{3.125}$Sn$_{0.875}$, and Mn$_{3.25}$Sn$_{0.75}$.** The left, middle, and right columns correspond to Mn$_3$Sn, Mn$_{3.125}$Sn$_{0.875}$, and Mn$_{3.25}$Sn$_{0.75}$. **a-c.** Electronic band structures along the high-symmetry path Γ-A-H-K-Γ of the primitive cell. The red solid lines and green dashed lines represent results from first-principles calculations and Wannier interpolation, respectively. **d-f.** AHC $\sigma_{(0001)}$ as a function of energy E. **g-i.** AHC $\sigma_{(01\bar{1}0)}$ as a function of energy E. All calculations include SOC and are performed within ±0.5 eV of the Fermi level. The arrows in d–i indicate the AHC values at the Fermi level.

## IV. SUMMARY

This work demonstrates that Mn enrichment in Mn$_3$Sn controllably induces a non-coplanar antiferromagnetic state, transforming the system from a coplanar 120° order. The transition is primarily mediated by four-spin ring exchange interaction in local triangular lattice instead of two-spin Heisenberg exchange interaction in global kagome lattice. With the help of the non-coplanar spin state, the time-reversal symmetry is broken and the finite $\sigma_{(0001)}$ is allowed. Both the anomalous Hall conductivity (AHC) from the (0001) basal plane $\sigma_{(0001)}$ and the $(01\bar{1}0)$ plane $\sigma_{(01\bar{1}0)}$ are calculated in stoichiometric and Mn-rich $Mn_3Sn$. A strong full-space AHE is expected in $Mn_{3.125}Sn_{0.875}$. The $\sigma_{(0001)}$ as high as -468 $\Omega^{-1}$ $cm^{-1}$ can be achieved, and the $\sigma_{(01\bar{1}0)}$ can be enhanced to -229 $\Omega^{-1}$ $cm^{-1}$ simultaneously. The realization of a full-space anomalous Hall response in an antiferromagnet represents a significant step toward three-dimensional topological spintronic functionality.

## ACKNOWLEDGMENTS

This work was supported by the National Natural Science Foundation of China (under grants 51901067, 51971087, 52101233, and 52071279), the Natural Science Foundation of Hebei Province (E2019205234), the Science and Technology Research Project of Hebei Higher Education (QN2019154), the Science Foundation of Hebei Normal University (L2019B11 and L2024B08), the Hebei Normal University Teaching Improvement Project (2022XJJG045), and the "333 Talent Project" of Hebei Province (C20231105).